\documentclass[12pt]{iopart} 

\usepackage{amssymb}

\def\beq{\begin{equation}}
\def\eeq{\end{equation}}
\def\beqn{\begin{eqnarray}}
\def\eeqn{\end{eqnarray}}
\newcommand{\nn}{\nonumber}
\newcommand{\wt}[1]{\widetilde{#1} }
 
\def\eqref#1{(\ref{#1})}

\begin{document}

\title{Analogue Gravity Models From Conformal Rescaling}

\author{Sabine Hossenfelder$^{1,2}$, Tobias Zingg$^2$} 

\address{$^1$ Frankfurt Institute for Advanced Studies\\
Ruth-Moufang-Str. 1,
D-60438 Frankfurt am Main, Germany
}

\address{$^2$ Nordita, Stockholm University and KTH Royal Institute of Technology\\
Roslagstullsbacken 23, SE-106 91 Stockholm, Sweden}

\eads{\mailto{hossi@fias.uni-frankfurt.de}, \mailto{zingg@nordita.org}}

\begin{abstract}

Analogue gravity is based on a mathematical identity between quantum field theory in curved space-time and the propagation of perturbations in certain condensed matter systems.
But not every curved space-time can be simulated in such a way. For analogue gravity to work, one needs not only a condensed matter system that generates the desired metric tensor, but this system then also has to obey its own equations of motion.
However, the relation to the metric tensor usually overdetermines the equations of the underlying condensed matter system, such that they in general cannot be fulfilled. In this case the desired metric does not have an analogue.

Here, we show that the class of metrics that have an analogue is larger than previously thought. The reason is that the analogue metric is only defined up to a choice of parametrization of the perturbation in the underlying condensed matter system.
In this way, the class of analogue gravity models can be vastly expanded.

\end{abstract}

\pacs{0440, 1110}






\submitto{\CQG}
\maketitle

\section{Introduction}
  
Some condensed matter systems act as `analogues' for gravity so that small perturbations
around their background fulfill an equation of motion formally identical to that of fields in a
curved space-time. The effective metric that quantifies the curved space-time is then a
function of the properties of the condensed matter system. 

There are various
examples for such analogies. For illustrative purposes we will deal here with the one that employs a non-viscous, irrotational, and barotropic fluid. The effective metric is then a function of the
pressure ($p_0$), the density ($\rho_0$), and the velocity field ($\vec v_0$) of the fluid -- though we note that the technique proposed here does not depend on the specific form of the metric and is more generally applicable. Here and
throughout this paper the index $0$ refers to the background field. 

That gravity and condensed matter physics are linked in this fashion has been known since the
mid 1980s~\cite{Unruh:1980cg, Barcelo:2005fc}, but only in the last
decade has the field begun to attract attention. One of the reasons is that new
connections between gravity and condensed matter systems have also been found
in different approaches, such as the AdS/CFT correspondence~\cite{Maldacena:1997re,Witten:1998qj,Gubser:2002tv}  and entropic gravity~\cite{Jacobson:1995ab,Padmanabhan:2009vy,Verlinde:2010hp,Verlinde:2016toy}.  The connection
between analog gravity and the AdS/{\sc CFT} duality was studied in 
\cite{Bilic:2014dda,Das:2010mk,holo,holo2,Chen:2012uc,Hossenfelder:2014gwa,Hossenfelder:2015pza}.
Another reason is that the possibility to experimentally explore these systems has recently 
become reality~\cite{Fedichev:2003id,Weinfurtner:2010nu,ex1,Steinhauer:2015saa,Euve:2015vml,Peloquin:2015rnl}.

This is an exciting research area because it allows to probe quantum field theory
in curved space-time, which is not currently possible in the actual curved space-time.
Beyond that, it allows to explore how the underlying, more fundamental, theory makes itself noticeable. For condensed matter systems the underlying theory
is known. For gravity it is not. So, from the correspondence between the two systems,
we might learn in which situations we can probe the underlying theory. 

But how much we can learn from analogue gravity is limited by the type of space-times that can be simulated in this fashion.
It is not generally the case that the fluid whose density, pressure, and velocity can be extracted from the metric fulfills the equations of motion. In fact,
for the most-studied case -- the Schwarzschild metric~\cite{Visser:1997ux,Garay:1999sk,Barcelo:2000tg}  -- the condensed matter
system does not fulfill the equation of motion.
This would lead one to conclude that it is not possible to simulate the
Schwarzschild metric directly, but only a metric which is conformal to
the Schwarzschild metric. We will, however, show here that by using a different way
to identify the background-perturbations, the Schwarzschild metric -- and many other metrics 
that were previously not possible -- can be simulated directly.

This can be achieved by observing that the way in which a conformal factor enters the equation of  motion of the perturbation is similar to the way that an effective mass-term does.
We hence modify the background system so that it creates an effective mass which cancels the
contribution of the conformal factor.
This method then allows us to find condensed matter systems that are gravitational analogues for space-times which are conformally equivalent to those space-times presently known to have
gravitational analogues.

This paper is organized as follows. In section~\ref{sec:acoustic_metric}, we briefly summarize the derivation of the equations of motion for the perturbations around the background
field, paying special attention to the effective mass and potential. In section~\ref{sec:rescaling},
we remind the reader how conformal rescaling affects the wave-equation.
In section~\ref{sec:examples} we show with several examples how an appropriately chosen conformal factor can be used to satisfy the equations of motion of the fluid describing a non-relativistic acoustic metric.
Then, in section~\ref{sec:finale}, we discuss the procedure in more general terms before summing up our
findings in the conclusions.

We use units in which the speed of light and $\hbar=1$. 
The constant $c$ denotes the speed of sound and {\sl not} the speed of light.
The metric signature is $(-1,1,1,1)$. 

\section{The Acoustic Metric}
\label{sec:acoustic_metric}

The effective metric which is perceived by small perturbations over a background field
can be derived using a Lagrangian approach. We here briefly summarize the results of this
derivation -- for details the reader is referred to~\cite{Barcelo:2005fc,Bilic:2013qpa,Hossenfelder:2014gwa}. 
We assume that the `subjacent' (as opposed to effective) metric lives on a Lorenzian manifold
of dimension $n+1$.

\subsection{Real Scalar Field}

We use a Lagrangian for a real scalar field which is of the general form
\beqn
{\cal L} = {\cal L}[\chi(\partial \theta), V(\theta,\vec x)] ~,
\eeqn
where 
\beqn
\chi = \eta^{\mu\nu} \left( \partial_\nu \theta \right) \left( \partial_\mu \theta \right)
\eeqn
is the kinetic term and $V(\theta,\vec x)$ describes the potential for the field and possible interactions with external sources.  Next, we split the field into a background and perturbations around the background, $\theta = \theta_0 + \theta_1$, where (by assumption) the background field $\theta_0$ fulfills the Euler--Lagrange equations. One can then expand the action in powers of the perturbation. This way, one finds that the terms for the perturbation take the form of the action of a scalar field propagating in an effective metric $g_{\mu\nu}$ which is defined by
\beqn
\sqrt{-g} g^{\mu \nu} = - \frac{\partial^2 {\cal L}}{\partial (\partial_\nu \theta) \partial (\partial_\mu \theta)} \Bigg|_{\theta=\theta_0} ~,
\label{eq:geff}
\eeqn
and has a mass term
\beqn
\sqrt{-g} m_{\rm eff}^2 = -  \Bigg[ \frac{\partial^2 {\cal L}}{\partial \theta \partial \theta} + \partial_\nu \left( \frac{\partial^2 {\cal L}}{\partial(\partial_\nu \theta) \partial \theta}\right) \Bigg]_{\theta=\theta_0}~.
\label{eq:meff}
\eeqn 
Since (\ref{eq:geff}) determines the propagation of sound-waves, it is also known as the (inverse) acoustic metric. 

The effective mass (\ref{eq:meff}) 
will not in general be a constant.  It might be more appropriate then to refer to it as potential. However, to avoid a nomenclature-confusion with the potential for the background field, we will (as it has become customary in the literature) refer to it it an effective mass. 

Next, one can represent the acoustic metric in terms of quantities familiarly used for fluid dynamics.
To that end, one introduces the stress-energy-tensor
\beqn
T_{\mu \nu} = (p_0+\rho_0)u_\mu u_\nu + p_0 \eta_{\mu \nu} ~,
\eeqn
where the four-velocity, pressure and density of the background field are given by
\beqn
u_\nu = \frac{\partial_\nu \theta}{\sqrt{\chi}}~,~p_0 = {\cal L}~,~\rho_0 = 2 \chi \frac{\partial {\cal L}}{\partial \chi} - {\cal L}~, \label{ident}
\eeqn
and the four-velocity is normalized to one
\beqn
\eta^{\mu\nu}u_\mu u_\nu = - 1~.
\eeqn
The field equations of the relativistic fluid are then identical to the conservation of the stress energy
\beqn
\partial_\nu T^{\mu \nu} = 0~,
\eeqn
and the acoustic metric and its inverse can be expressed as
\beqn
g^{\mu \nu} &=& c^{\frac{2}{n-1}}  \left(  \frac{\rho_0+p_0}{\chi} \right)^{-\frac{2}{n-1}}  \left( \eta^{\mu\nu} + \left(1- \frac{1}{c^2} \right) u^\mu u^\nu \right) ~, \\
g_{\mu \nu} &=& c^{\frac{-2}{n-1}}  \left(  \frac{\rho_0+p_0}{\chi} \right)^{\frac{2}{n-1}}  \left( \eta_{\mu\nu} + \left(1- c^2 \right) u_\mu u_\nu \right) ~. \label{endderiv}
\eeqn
Here, $c$ is the speed of sound and defined by $c^{-2} = \partial \rho_0/\partial p_0$.

In the non-relativistic limit, $p_0 \ll \rho_0$ and $v^2 \ll c^2$, then the acoustic metric is of the form
\beqn
g^{\mu \nu} (t, {\vec x}) &\propto& \left( \frac{\rho_0}{c} \right)^{-\frac{2}{n-1}}
\left( \begin{array}{cc}
-1/c^2 & - v_0^j/c^2 \\
-v_0^i/c^2 & \delta^{ij} - v_0^i v_0^j/c^2  \end{array} \right)~, \label{gupd} \\
g_{\mu \nu} (t, {\vec x}) &\propto& \left( \frac{\rho_0}{c} \right)^{\frac{2}{n-1}}
\left( \begin{array}{cc}
-(c^2-v_0^2) & - (v_0)_j \\
- (v_0)_i & \delta_{ij}   \end{array} \right) \label{gdownd} ~.
\eeqn
In this limit, the equations of motion for the background field are the continuity equation and the 
Euler equation:
\beqn
\partial_t \rho_0 + {\vec \nabla} \cdot (\rho_0 {\vec v}_0 ) &=& 0 \label{continuity} ~, \\
\rho \left[ \partial_t {\vec v}_0 + ( {\vec v}_0 \cdot {\vec \nabla})  {\vec v}_0 \right] &=& \vec{F} \label{euler}~.
\eeqn
The argument we will develop here assumes that we are in the non-relativist limit.

We will also assume, as usual, that the fluid is non-viscuos, has vanishing rotation (i.e.~is vorticity-free), and is barotropic. 
The velocity field is then the gradient of a scalar field ${\vec v}_0 = - {\vec \nabla} \phi$ and the density $\rho_0$ is a function of $p_0$ only. 
In this case, the Euler equation can be integrated once and can be written as
\beqn
 \partial_t \phi =  h + \frac{1}{2} \left( \vec \nabla \phi \right)^2~, \label{euler3}
\eeqn
where 
\beqn
h(p) = \int_0^p \frac{dp'}{\rho_0(p')} ~
\eeqn
is the specific enthalpy.

Let us then make the following observation. Consider we have a type of fluid with a specified
Lagrangian $\cal{L}$ and an unspecified potential $V$. We would like to find a potential
for the fluid that
realizes a space-time metric which takes the form of the acoustic metric (\ref{gdownd}). 
We then first read off $\rho$ and $\vec v$ from the metric. With the previously made assumptions that the fluid be non-viscuos, vorticity-free and barotropic, this will result in two equations of motions (\ref{continuity}) and (\ref{euler3}).
In general, however, both equations cannot be solved simultaneously. And since~\eqref{eq:geff} does not depend on $V$ directly,  different choices of the potential do not remedy the problem. This leads to the conclusion that most metrics cannot be realized as gravitational analogues. Indeed this is the case e.g.~for the Schwarzschild-metric 
\cite{Barcelo:2005fc,Hossenfelder:2014gwa}. 

We will in the following show how to circumvent this impasse.

\subsection{Complex Scalar Field}

Before we get to our main argument, let us briefly look at how to generalize the
above formalism to a complex scalar field, when the Lagrangian is a
function of $\partial \theta,\partial \theta^{*},$ and $\theta,\theta^{*}$.
The expansion works similarly to the case of a real scalar. When we 
set 
\beqn
\Theta_{0/1} = \left( \theta_{0/1}, \theta_{0/1}^{*} \right) 
\eeqn
 for brevity, then
expanding the action to second order
leads to the equations of motion for the perturbations
\beqn
\partial_\mu \left( G^{\mu\nu} \cdot \partial_\nu\Theta_1 \right) - M \cdot \Theta_1  = 0
\, ,
\eeqn
where
\beqn
G^{\mu\nu} &=&
\left[\begin{array}{cc}
\frac{1}{2}\left( \frac{\partial^2 \mathcal{L}}{\partial (\partial_\mu \theta) \partial\theta^{*}_{;\nu}} + c.c. \right)	& \frac{\partial^2 \mathcal{L}}{\partial\theta^{*}_{;\mu}\partial\theta^{*}_{;\nu}}	\\
\vspace{-2mm}  & \\
\frac{\partial^2 \mathcal{L}}{\partial\theta^{}_{;\mu}\partial\theta^{}_{;\nu}}	&  \frac{1}{2}\left( \frac{\partial^2 \mathcal{L}}{\partial\theta_{;\mu}\partial\theta^{*}_{;\nu}} + c.c. \right)
\end{array}\right]
\, , \\
M &=&
\left[\begin{array}{cc}
\frac{1}{2}\left( \frac{\partial^2 \mathcal{L}}{\partial\theta \partial\theta^{*}} - \partial_\mu\frac{\partial^2 \mathcal{L}}{\partial\theta \partial\theta^{*}_{;\mu}} + c.c. \right)	& \frac{\partial^2 \mathcal{L}}{\partial\theta^{*} \partial\theta^{*}} - \partial_\mu\frac{\partial^2 \mathcal{L}}{\partial\theta^{*} \partial\theta^{*}_{;\mu}}	\\
\vspace{-2mm}  & \\
\frac{\partial^2 \mathcal{L}}{\partial\theta^{}\partial\theta^{}} - \partial_\mu\frac{\partial^2 \mathcal{L}}{\partial\theta \partial\theta^{}_{;\mu}}	&  \frac{1}{2}\left( \frac{\partial^2 \mathcal{L}}{\partial\theta \partial\theta^{*}} - \partial_\mu\frac{\partial^2 \mathcal{L}}{\partial\theta \partial\theta^{*}_{;\mu}} + c.c. \right)
\end{array}\right]
\, .\qquad
\eeqn
Here, a semicolon denotes a covariant derivative with respect to the index that follows. Assuming $\mathcal{L}$ is as usual real-valued, these matrices are Hermitian. 

The equation of motion for the perturbations $\theta_1$ and $\theta_1^{*}$ are then two separate equations, which are complex conjugates of each other.
They do not in general split into separate equations for $\theta_1$ and $\theta^{*}_1$ respectively, which means there is no straight-forward
interpretation of the equations of motion in terms of an analogue 
metric.

However, under certain circumstances it is possible to separate the two equations.
In particular, if $G^{\mu\nu}$ and $M$ have a common eigenvector, then a perturbation which is parallel to this eigenvector will separate
from an orthogonal vector. 
Such a common eigenvector of $G_{\mu\nu}$ and $M$ exists when the condition is fulfilled that
\beqn
\ker \bigcap\limits_{k,l=1}^{2} \left.\left[ G_{\mu\nu}^k, M^l \right]\, \right|_{\Theta = \Theta_0} \neq \{0\}
\, .
\eeqn
Here, $k,l$ are exponents, not indices, and the square brackets denote the commutator. 
We will in the following not explore the necessary conditions which this implies for the Lagrangian. We merely
note that a sufficient condition to fulfill this requirement is that $\mathcal{L}$ is subject to a reality condition, and background as well 
as perturbation are restricted to real values. We will in the following assume that
this is the case. If the original definition of the fields does not fulfill this condition, then one can separate 
the equations by forming the linear combinations $\theta_1+\theta_1^*$ and $\theta_1-\theta_1^*$. In the case of a $U(1)$ gauge symmetry, this does not even impose any actual restriction, as a choice orthogonal to the real solution would be pure gauge.

\section{Conformal rescaling of the wave-equation}
\label{sec:rescaling}

The perturbation of the scalar field, $\theta_1$, satisfies the equation of motion
\beqn
\square \theta_1 - m_{\rm eff}^2 \theta_1
	= \frac{1}{\sqrt{|g|}} \partial_\mu \left(\sqrt{|g|} g^{\mu\nu} \partial_\nu \theta_1 \right) -  m_{\rm eff}^2 \theta_1
	= 0 ~.
\label{eq:KG_01}
\eeqn
We now introduce a rescaled metric $\wt{{\bf g}} = \Omega^{-2} \bf{g}$, with an unspecified conformal factor $\Omega(t, \vec x)$ that is a function of the coordinates.
The d'Alembert-Operator $\square$ acting on $\theta_1$ can then be rewritten via replacing occurrences of $\bf{g}$ in terms of $\wt{{\bf g}}$ and $\Omega$,
\beqn
\square \theta_1
	&=& \frac{1}{\Omega^n\sqrt{|\wt{g}|}} \partial_\mu \left(\Omega^{n-2}\sqrt{|\wt{g}|} \wt{g}^{\mu\nu} \partial_\nu \theta_1 \right)
\ .
\eeqn
Next, we  replace $\theta_1$ with a rescaled field $\wt{\theta_1} = \Omega^{\frac{n-2}{2}} \theta_1$.
This results in the identity
\beqn
\square \theta_1
	&=&	\frac{1}{\Omega^n\sqrt{|\wt{g}|}} \partial_\mu \left[
				\sqrt{|\wt{g}|} \wt{g}^{\mu\nu}\left(
					\Omega^{\frac{n-2}{2}} \partial_\nu \wt{\theta}_1
					+ \frac{2-n}{2}\Omega^{\frac{n-4}{2}} \wt{\theta}_1 \partial_\nu \Omega
					 \right)\right]	\nn\\
	&=&	\frac{\Omega^{-\frac{n+2}{2}}}{\sqrt{|\wt{g}|}} \partial_\mu \left( \sqrt{|\wt{g}|} \wt{g}^{\mu\nu} \partial_\nu \wt{\theta}_1 \right)
		+ \frac{(2-n) \wt{\theta}_1}{2 \Omega^n\sqrt{|\wt{g}|}} \partial_\mu \left( \Omega^{\frac{n-4}{2}} \sqrt{|\wt{g}|} \wt{g}^{\mu\nu} \partial_\nu \Omega \right)
\ .
\eeqn
Plugging this result into the original equation of motion, we therefore see that any field $\theta_1$ which fulfills the wave-equation~\eqref{eq:KG_01} in the background given by the metric ${\bf g}$ can 
be mapped to a field $\wt \theta_1$ which fulfills the wave-equation in the background $\wt{\bf g}$
\beqn
\wt{\square} \wt{\theta}_1 - \wt{m}_{\rm eff}^2 \wt{\theta}_1
	= 0
\ ,
\label{eq:KG_02}
\eeqn
provided we define the new effective mass as
\beqn
\wt{m}_{\rm eff}^2 = \Omega^2 {m}_{\rm eff}^2 + \Omega^{\frac{2-n}{2}} \wt{\square} \Omega^{\frac{n-2}{2}} 
\ .
\label{eq:V_trafo}
\eeqn
The conformal factor hence determines which effective mass is necessary to obtain the wave-equation in the desired background ${\bf g}$.

 This also works if the scalar field is U(1)-charged and has a gauge field $A_\nu$. We then have
\beqn
(\nabla_\mu+iqA_\mu)(\nabla^\mu+iqA^\mu) \theta_1 - {m}_{\rm eff}^2 \theta_1
	= 0
\ ,
\label{eq:cKG_01}
\eeqn
and the equation can be mapped to 
\beqn
(\wt{\nabla}_\mu+iq A_\mu)(\wt{\nabla}^\mu+iq A^\mu) \wt{\theta}_1 - \wt{m}_{\rm eff}^2 \wt{\theta}_1
	= 0
\ ,
\label{eq:cKG_02}
\eeqn
with the same effective mass as in Equation \eqref{eq:V_trafo}. Here, $\nabla$ and $\wt \nabla$ denote the covariant
derivatives compatible with the effective metrics ${\bf g}$ and $\wt {\bf g}$, respectively. The gauge field remains unmodified.

\section{Examples}
\label{sec:examples}

Space-times that can be simulated with gravitational analogues
include the Schwarzschild black hole~\cite{Visser:1997ux,Garay:1999sk,Barcelo:2000tg} and expanding de-Sitter space that mimics
the inflationary epoch of the early universe~\cite{Volovik:2000ua,Weinfurtner:2004mu,Jain:2007gg,Lin:2012ft,Bilic:2013qpa}. 

In many cases, it was found that the specific metric of interest does not satisfy the resulting fluid's equations of motion, only a metric conformally equivalent to it does.
To study certain phenomena from a qualitative point of view, this is sufficient, but if any scale-dependence is to be taken seriously in such findings, it would be much preferable to directly simulate the original metric and not one that is merely conformal to it.

Important examples of metrics that can be made to satisfy the equations of motion by introducing a conformal factor are below.

\subsection{Conformally Flat Space-times}

The simplest case to illustrate the use of what we have shown in the previous section is to consider a conformally flat space-time with line-element
\beqn
{\rm d} s^2 = a^2(t,\vec x) \left( - {\rm d}t'^2 +  \delta_{ij} {\rm d} x^i {\rm d} x^j  \right) ~. \label{cflat}
\eeqn
Formally, this case may not seem particularly interesting, because any conformally flat metric with $\partial_t \rho_0 = 0$ can directly be realized as an acoustic metric by choosing $\rho_0 = a^2$ and $\vec v_0 = 0$.
Nevertheless, for an arbitrary function $a(t, \vec x)$, it may not be feasible to create an experimental setup where the density $\rho_0$ has a complicated, spatially modulated, and potentially time-dependent, profile.
But with the derivation from section~\ref{sec:rescaling}, where is was laid out that the analog metric is actually only determined up to a choice of parametrization, by setting $\Omega = a$, a perturbation in~\eqref{cflat} can be mapped to a perturbation propagating in flat Minkowski space.
This, of course, comes at the price of having to adjust the effective mass~\eqref{eq:V_trafo} for the perturbation to accommodate for the shift due to the conformal factor.
From the point of view of an experimental setup, however, the latter would appear much more practical, e.g.~by coupling to an external potential, which seems more straightforward to realize than having to change the background altogether to allow for different profiles $a(t,\vec x)$.

\subsection{Black Holes}
\label{sec:BH}

A more interesting case are black hole space-times, which, in general, are not conformally flat -- e.g.~the Schwarzschild metric. They are of special interest as their analogue dual can be used to test the presence of Hawking radiation.\footnote{
see e.g.~\cite{Fedichev:2003id,Weinfurtner:2010nu,Steinhauer:2015saa,Euve:2015vml}}

Thus, consider a typical static stationary black hole space-time in $n+1$ dimensions\footnote{Schwarzschild, AdS-Schwarzschild, RN and AdS-RN can all be written in a form like \eqref{eq:BH_st}}
\beqn
{\rm d} s^2 = \Omega(r)^2 \left[ - \gamma(r) {\rm d}t^2 + \frac{{\rm d} r^2 }{\gamma(r)} + r^2 {\rm d} \sigma^{n-1} \right]~, 
\label{eq:BH_st}
\eeqn
with the horizon topology of a $n-1$ dimensional sphere ($\rm{d} \sigma$) and blackening factor $\gamma(r)$.
In analogy to Painlev\'e--Gullstrand coordinates for the Schwarzschild metric~\cite{PG1,PG2,PG3} this metric can be brought to the form
\beqn
{\rm d} s^2 = \Omega(r)^2 \left[ - \kappa^2\,\gamma(r) {\rm d}t'^2 + 2\kappa\sqrt{1-\gamma(r)}\,{\rm d} t' {\rm d}r + {\rm d}r^2 + r^2 {\rm d} \sigma^{n-1} \right]~,
\label{eq:BH_st_PG}
\eeqn
where the constant $\kappa$ has been introduced for later convenience.
Comparing to \eqref{gdownd} it is now straightforward to read off the fluid components,
\beqn
c_0 = \kappa~,\quad
\rho_0 = \kappa\,\Omega(r)^{n-1}~,\quad
v_0^r = \kappa \sqrt{1-\gamma(r)}~.
\label{eq:BH_fluid}
\eeqn
For generic $\gamma$ and $\Omega$ it can easily be checked that the continuity equation~\eqref{continuity} is generally not satisfied, unless the very specific condition $\rho_0 v^r_0 \sim r^{1-n}$ is met.
However, if $\Omega(r)$ were an adjustable function, the continuity and Euler equation could be solved by choosing
\beqn
\Omega(r) = \frac{1}{r}\left[ 1 - \gamma(r) \right]^{1/(n-1)}~,\quad
F^r = - \frac{\kappa^3}{r^{n-1}} \frac{\gamma'(r)}{\sqrt{1-\gamma(r)}}~.
\label{eq:BH_OF}
\eeqn
We can therefore conclude that while a generic black hole space-time will likely not be an analogue metric, there is usually a conformally related metric that actually will be. As per our previous argument, this means that we can find a rescaled perturbation for which the black hole space-time is the analogue metric. 

Note, however, that this may not work globally, as e.g.~in regions with $\gamma(r) > 1$ bringing the metric directly to Painlev\'e--Gullstrand form is not well defined -- though, in areas close enough to the horizon, there are no such issues.

\subsection{Black Branes}

With black brane we refer to a space-time with a planar event horizon.
A typical metric is of the form
\beqn
{\rm d} s^2 = \Omega(r)^2 \left[ - \gamma(r) {\rm d}t^2 + \frac{{\rm d} r^2 }{\gamma(r)} + {\rm d} {\bf x}^2 \right]~.
\label{eq:BB_st}
\eeqn
Such space-times are of particular relevance in AdS/{\sc CFT}, where one is often interested in a gravitational dual of a strongly correlated thermal field theory on the flat boundary geometry in aymptotically AdS space-times.
Finding an analog dual of such a metric works almost exactly as in the previous section.
From transforming into Painlev\'e--Gullstrand type coordinates,
\beqn
{\rm d} s^2 = \Omega(r)^2 \left[ - \kappa^2\,\gamma(r) {\rm d}t'^2 + 2\kappa\sqrt{1-\gamma(r)}\,{\rm d} t' {\rm d}r + {\rm d}r^2 + {\rm d} {\bf x}^2 \right]~,
\label{eq:BB_st_PG}
\eeqn
the fluid components can again be read off directly.
The main difference to the previous case is that the spatial part of the line element is now in Cartesian rather than spherical coordinates.
This slightly changes how $\Omega(r)$ needs to be chosen to satisfy the fluid equations.
A quick calculation reveals that the fluid components are again as in~\eqref{eq:BH_fluid}, but what has changed is
\beqn
\Omega(r) = \left[ 1 - \gamma(r) \right]^{1/(n-1)}~,\quad
F^r = - \kappa^3 \frac{\gamma'(r)}{\sqrt{1-\gamma(r)}}~.
\label{eq:BB_OF}
\eeqn

\section{Analogue Systems from Conformal Rescaling}
\label{sec:finale}

The cases discussed in section~\ref{sec:examples} exemplify a more general lesson which can be summarized as follows.

For a generic analougue metric of the form~\eqref{gdownd}, the fluid equations are usually not satisfied unless special compatibility conditions are met.
The introduction of a conformal factor allows to resolve that obstacle because it adds an additional degree of freedom. With the conformal factor taken into
account, the continuity and Euler equation~(\ref{continuity},\ref{euler}) are not any more overdetermined and it is now generally possible to
realize a given background as an analogue metric with an appropriately chosen $\Omega$.

As previously mentioned, a caveat is that the solution might not be valid everywhere as, e.g.~for a generic background and boundary conditions it is a priori not necessarily forbidden that~$\Omega$ could potentially change sign, which would result in an unphysical analogue metric.
However, locally the system is solvable and in case $\Omega$ would change sign this just means that the parameter range would have to be reduced in a conceivable experimental setup.

We can then combine this insight with what we showed in section~\ref{sec:rescaling}.
There, we laid out that the analogue metric is actually only determined up to a choice of parametrization of the perturbation.
In particular, a conformal rescaling of the analogue metric is related to a conformal rescaling of the perturbation and a shift in the potential that determines the effective mass.
The latter is of course more than a mere choice of parametrization; it is a change to the experimental setup.
Nevertheless, it is still an approach that seems experimentally quite conceivable to realize, e.g.~by coupling to an external potential.

Hence, we can conclude that any given metric of the form of a non-relativistic acoustic metric~\eqref{gdownd} can indeed be realized as an analogue metric -- albeit with some restrictions on global validity as pointed out above.
The procedure, as outlined above, is to first use the liberty to choose a representative of the conformal class of metrics that satisfies the fluid equations, and then absorb the conformal factor into a change of the potential for the perturbation.

We have in this present work not investigated the relativistic case. It is more complicated because the pressure appears in the effective metric and the equations of motion cannot as easily be integrated once. It is therefore not a priori clear
that introducing one additional free function will in general allow to solve the equations of motion. It is clear, however,
that the additional function will also in this case increase the class of metrics that can be realized. However, just in which way the class would be enlarged 
is beyond the scope of this present work.

\subsection{Modified Lagrangian}
\label{sec:lagrangian}

Having established how a conformal factor in the analogue metric can be absorbed into the effective mass of the perturbation, the question becomes how this change can be incorporated into the Lagrangian.
Ideally, one would wish to change the potential for the scalar perturbation without having to change anything in the background metric, or the fluid, respectively.

Thus, assume there are two actions $S_{1,2}[\theta]$ such that, for a particular solution $\theta=\theta_0$, in either case the Euler--Lagrange equation is satisfied, as well as the resulting analog metric and the stress tensor being identical.
Then, on general grounds, it can be assumed that, when evaluated close to the particular solution,
\beqn
S_1[\theta] - S_2[\theta] = \mathcal{O}(|\theta-\theta_0|^2)~.
\eeqn
This constrains the modified Lagrangian in the following way.
If ${\mathcal{L}}[\partial_\nu\theta,\theta]$ is given and the Euler--Lagrange equation is solved at $\theta = \theta_0$, then any Lagrangian that reproduces the same background and analog metric is expected to be of the form $\tilde{\mathcal{L}}[\partial_\nu\theta,\theta,\Delta]$ with $\Delta = (\theta-\theta_0)^2$ such that $\tilde{\mathcal{L}}[\partial_\nu\theta,\theta,0] = {\mathcal{L}}[\partial_\nu\theta,\theta]$.
Of course, under the condition that $\tilde{\mathcal{L}}$ is sufficiently smooth in the third variable, such that the resulting Euler--Lagrange equation are still well-defined and fulfilled at~$\theta=\theta_0$.

Then, it is straightforward to verify that analogue metric~\eqref{eq:geff} is unchanged and the fluid identification, as well as the stress-energy tensor in this particular background remain the same.
However, according to~\eqref{eq:meff}, the effective mass for the variation will change to
\beqn
\sqrt{-g}\tilde{m}_{{\rm eff}}^2 &=& \sqrt{-g}{m}_{{\rm eff}}^2 - 2 \frac{\partial \mathcal{L}}{\partial \Delta}
~.
\label{eq:mass-change}
\eeqn
The question remains what practical ways there are to modify a given Lagrangian.
This, however, is a very model-specific problem which cannot be answered in all generality.
 
\section{Conclusion}
\label{sec:blabla}

In this work we have detailed how an arbitrarily adjustable change in parametrization of a perturbation around a condensed matter background changes the resulting analogue metric and potential.
This results in a significant extension of the class of space-times that have a fluid analogue.

Using this freedom in parametrization, we considered metrics -- in particular black hole space-times -- which were hitherto argued to only be realizable in a conformally equivalent way as analogue metrics. We then showed how any introduced conformal factor could be absorbed into a change of potential for the perturbation.
By using this procedure, we concluded that, once a specific space-time has been specified, having to fulfill the equations of motion for the analogue condensed matter system is less restrictive than it might originally have appeared.
In the case of non-viscous, barotropic fluids in the non-relativistic limit, we even found that this new degree of freedom is sufficient to prevent the fluid equations of motion from becoming overdetermined, thus removing a feature that otherwise would obstruct finding a consistent analogue fluid.

We wish to emphasize that the re-parametrization introduced here does not rely on specific symmetries of an analogue model, but merely quantifies a freedom of choice for selecting the perturbation and defining its analogue metric.
Therefore, the procedure as outlined in this paper can be applied to any conceivable analogue model and represents a powerful new tool to extend the classes of analogue metrics.

\section*{Acknowledgements}

TZ acknowledges support from Vetenskapsradet under the project number 2015-04852. SH acknowledges support by the Foundational Questions Institute (FQXi).

\end{document}